\documentclass[reprint,aps,prstper]{revtex4-1}

\usepackage{graphicx}
\usepackage{hyperref}
\usepackage{footnote}
\usepackage{amsmath,amssymb}

\begin{document}

\title{Framework of goals for writing in physics lab classes}

\author{Jessica R. Hoehn}
\email[]{jessica.hoehn@colorado.edu}
\affiliation{Department of Physics, University of Colorado, 390 UCB, Boulder, Colorado 80309, USA}
\affiliation{JILA, National Institute of Standards and Technology and University of Colorado, Boulder, Colorado 80309, USA}

\author{H. J. Lewandowski}
\affiliation{Department of Physics, University of Colorado, 390 UCB, Boulder, Colorado 80309, USA}
\affiliation{JILA, National Institute of Standards and Technology and University of Colorado, Boulder, Colorado 80309, USA}

\begin{abstract}
Writing is an integral part of the process of science. In the undergraduate physics curriculum, the most common place that students engage with scientific writing is in lab classes, typically through lab notebooks, reports, and proposals. There has not been much research on why and how we include writing in physics lab classes, and instructors may incorporate writing for a variety of reasons. Through a broader study of multiweek projects in advanced lab classes, we have developed a framework for thinking about and understanding the role of writing in lab classes. This framework defines and describes the breadth of goals for incorporating writing in lab classes, and is a tool we can use to begin to understand \textit{why}, and subsequently \textit{how}, we teach scientific writing in physics.
\end{abstract}

\maketitle

\section{\label{sec:intro}Introduction}
Laboratory classes are an essential element of the undergraduate physics curriculum; they afford opportunities for students to learn lab skills such as troubleshooting or modeling, learn physics content, understand how the community of practicing physicists engages in the process of experimentation, and develop an identity as a physicist~\cite{Dounas-FrazerLewandowski2016, HolmesEtal2017, HolmesWieman2018, IrvingSayre2014, MoskovitzKellogg2011, Dounas-FrazerLewandowski2018, Lerner2007}. As such, physics education researchers are increasingly attending to investigations and improvements of student learning in lab class environments. 

A key element of most lab classes is scientific documentation and writing in the form of lab notebooks and lab reports. Writing plays a central role in the doing and learning of science, and is identified by the Joint Task Force on Undergraduate Physics Programs (JTUPP) as one of the skills that students need to develop in order to be successful in a wide variety of careers upon receiving a physics bachelors degree~\cite{JTUPP2016}. More specifically, the American Association of Physics Teachers (AAPT) published a set of recommendations for the undergraduate physics laboratory curriculum that includes ``communicating physics'' as a key learning outcome that lab classes should attend to~\cite{KozminskiEtal2014}. Some argue that the \textit{best} place to teach scientific writing skills is in lab classes, where students actually \textit{do} physics~\cite{MoskovitzKellogg2011}. Whether it is because most of the ``doing'' of science takes place in lab classes, or because lab classes provide the most flexibility in terms of time and content coverage, most of the writing that physics students encounter in their undergraduate career takes place in lab classes.  

Within the realm of laboratory instruction, project-based labs are gaining traction. In a \textit{Physics Today} article, Feder describes the recent trend and says that ``Implementation varies, but the crosscutting aims are to motivate students and prepare them for the changing needs of the modern world and workplace. Learning by doing is a common thread. In the process, students are to develop into team players and effective communicators''~\cite[p.28]{Feder2017}. Additionally, JTUPP suggests that student-designed projects in advanced lab courses have the benefit of providing ``authentic research and communication experiences''~\cite{JTUPP2016}; they recommend that advanced lab courses implement some form of multiweek research projects in order to prepare students for $21^{st}$ century careers. The scientific writing associated with project-based labs may include proposals, lab notebooks, and final reports in the style of a journal article. An overarching goal of advanced project-based labs is to help students become more central members of the community of practicing physicists~\cite{IrvingSayre2014}, and scientific writing may serve as one element through which this goal can be achieved. Thus, the specific goals we have for writing in lab classes will be situated within this broader context. In an ethnographic case study of a physics graduate student learning to write a journal article in situ, Blakeslee suggests that a general goal for students taking up scientific writing practices is to help students transition from a traditional scaffolded educational environment like a lab class or guided research experience to a more independent and open ended scenario where they are responsible for producing and disseminating knowledge~\cite{Blakeslee1997}. In this paper, we identify possible goals for incorporating writing in physics lab classes, noting that advanced lab classes (especially those with a student-designed multiweek final project) may play a unique role in preparing students for future research experiences. This is true for many aspects of experimental physics training, including engaging in scientific writing.

While there is a growing body of research on teaching and learning in physics lab classes, there is little published on the specific role that writing can (or should) play in project-based labs. We begin to address this gap by asking the following research question: \textit{Why might instructors incorporate writing during multiweek final projects in advanced labs?} Once we understand the goals for writing in project-based lab courses, we can begin to investigate \textit{how} writing can best be implemented to achieve such goals. To that end, the purpose of this paper is to present a framework for thinking about and understanding the role of writing in physics labs. We interviewed four advanced lab instructors and conducted a coding analysis to identify their goals for incorporating writing in student-designed final projects. We then supplemented the results of the coding analysis with common ideas from the literature in order to develop a framework that identifies various reasons physics lab instructors \textit{might} assign writing, thus creating a framework that is broadly applicable to physics lab classes. 

As a general structure or scaffolding for information, a framework can serve a variety of purposes. In physics education research, frameworks are commonly used as tools for both research and teaching (e.g., ~\cite{ZwicklEtal2015, Dounas-FrazerLewandowski2018, HoehnFinkelstein2018, PawlakIrvingCaballero2018, Hyater-AdamsEtal2018, Chi2009, MarshmanSingh2015, HammerEtal2005,WilcoxEtal2013}); a framework may be used to: a) understand or analyze a phenomenon or topic of interest, b) inform or identify future research questions, c) inform pedagogical decisions or curricular design, or d) advance theory development in a given sub-domain. Our framework, which consists of fifteen possible goals for writing in physics lab classes organized into five categories, has potential to serve the first three purposes. In this paper, we present the framework, providing examples from data and literature in which it is grounded. In Section \ref{sec:implications}, we discuss the implications for research and teaching and call for more research in the arena of writing in physics labs.   

\section{\label{sec:background}Background}
There is a considerable amount of literature on teaching writing in general, and the role of writing in science specifically. However, very little of this work has been employed in the specific domain of physics lab classes. In this section, we provide a brief overview of the literature on writing reforms and how they have been realized in science (Section \ref{sec:background-writing}), and then we outline a few specific approaches to writing in lab classes in physics or other related fields (Section \ref{sec:labwriting}). 

\subsection{\label{sec:background-writing}Approaches to teaching writing}
Writing across the curriculum (WAC), also sometimes referred to as writing in the disciplines (WID), is a movement within composition studies that took hold in the US in the early $20^{th}$ century and by the 1970s was beginning to be common across higher education institutions~\cite{WACHistory2005}. The goal of the WAC movement was to distribute writing instruction across the entire undergraduate curriculum, implementing writing within each discipline rather than relegating it to \textit{only} english, composition, or literature courses. This was a response, in part, to ``the recognition that different disciplines are characterized by distinct ways of writing and knowing''~\cite{Carter2007}. 

Many universities have centralized writing centers that serve as the focal point for writing on campus, providing resources to individual departments implementing writing in some way~\cite{WACHistory2005}. As such, today it is common place for students to encounter some form of writing in many of their courses, regardless of major or discipline. However, the extent to which writing has been implemented in the disciplines has varied. Lerner suggests that in science courses, writing has often been implemented in superficial ways, focusing on grammar and formatting rather than the deeper more fundamental writing skills like argument construction or organization of ideas~\cite{Lerner2007}. This lackluster implementation also goes the other way around---in the context of first year writing courses that focus specifically on science or scientific writing, Moskovitz and Kellogg argue for the inclusion of ``primary science communication'' (i.e., journal articles) because learning to write scientifically should involve reading actual scientific writing~\cite{MoskovitzKellogg2005}. 

Within the literature on writing, there are different paradigms or views of the nature of writing. Perhaps the most salient and intuitive approach for physicists and physics educators is the idea of writing as communication. Communication is fundamental to the creation and advancement of scientific knowledge, and writing is the primary medium through which that takes place. We need to communicate our theories, models, results, and conclusions clearly and effectively to other scientists as well as the general public. The idea of writing as communication foregrounds the final product of a given piece of writing, which serves to demonstrate what the scientist (or science student) knows, or what they accomplished. 

In contrast, a Writing to learn (WTL) approach focuses on the \textit{process} of writing as a tool for thinking and learning. Many composition scholars suggest that too much emphasis is placed on writing as communication and not enough on ``writing as articulation''~\cite{Rivard1994} or ``thinking on paper''~\cite{Howard1988}. Writing is a messy and iterative process that can be used to construct knowledge or understanding, clarify ideas, generate a personal response to a phenomenon or subject, figure out solutions to complex problems, construct and critique arguments, synthesize ideas, or reflect on your own knowledge~\cite{Rivard1994, ReynoldsEtal2012}. Rather than viewing writing as packaging for already formulated ideas, the WTL approach emphasizes the act of formulating those ideas. Bean suggests that part of teaching students about this process of writing means teaching them to revise, and that the process of revision takes writing from being ``writer-based'' (the stage when a writer is clarifying meaning for themselves) to being “reader-based” (when the writer is focused on clarity for the audience)~\cite{Bean2011}. The WTL approach has been integrated into STEM classes through writing assignments that explicitly guide students through the processes of reflection or argument construction~\cite{ReynoldsEtal2012, Eblen-Zayas2016, Dounas-FrazerReinholz2015, MoskovitzKellogg2011}. 

Shifting from a writing as communication to WTL approach is akin to shifting from a ``knowledge telling'' to ``knowledge generating'' epistemology~\cite{Bean2011, ReynoldsEtal2012}. That is, these different approaches to writing are fundamentally connected to different views about the nature of knowledge. A view of knowledge as discrete pieces of information to be studied or memorized lends to a view of writing as information or demonstration of “correct” facts, while a view of writing as argument and analysis is aligned with a view of knowledge as dialogic, contingent, ambiguous, or tentative~\cite{Blakeslee1997}. This matters because the way we, as teachers, view writing impacts how we implement it in our classes and thus how students come to view writing~\cite{ReynoldsEtal2012}. Composition scholars suggest that students' prose will be “cognitively immature” if they see knowledge as simply acquisition of correct information~\cite{Bean2011}. However, this does not mean we should abandon the idea of writing as communication; clear and effective communication is crucial for the advancement of science. Often, the writing as communication and WTL approaches are in tension with one another (as illustrated by the fact that most definitions of WTL describe it \textit{in contrast} to communication), but they need not be. Bean describes writing as ``both a process of doing critical thinking and a product communicating the results of critical thinking''~\cite{Bean2011}; we need both aspects, and can attend to them simultaneously. 

In STEM disciplines specifically, a third approach---Writing as professionalization (WAP)---has been employed to emphasize the fact that writing can be a way to master disciplinary forms of reasoning or argumentation, and can facilitate learning content and ways of thinking that are specific to a discipline~\cite{ReynoldsEtal2012, Carter2007}. The process of learning professional discourse norms and experiencing the central role of written communication in the process of science can help students become more central members of the community of practicing physicists~\cite{QuanElby2016, IrvingSayre2014, LaveWenger1991}. This idea of WAP is prevalent in STEM, and specifically in lab classes where students most often encounter writing~\cite{MoskovitzKellogg2011}. The AAPT recommendations for undergraduate physics laboratory curricula align with the WAP approach by suggesting that students should learn to communicate ``in forms authentic to the discipline,'' communicate arguments using appropriate technical vocabulary, present data with appropriate significant figures and uncertainty, and present data, ideas, or results in appropriately professional plots, tables, diagrams, and schematics~\cite{KozminskiEtal2014}. 

Moskovitz and Kellogg indicate that most recent reforms to writing in lab classes have taken \textit{either} a WAP or a WTL approach, identifying a tension between the two~\cite{MoskovitzKellogg2011}. They suggest that when writing is solely used as a means to an end, it is ``likely to be at odds with [expectations] of professional scientific discourse.'' Our interpretation of the literature we have drawn from in this section is that each of the three approaches to writing---Communication, WTL, and WAP---are necessary for writing in science, and that we can attend to all three simultaneously, perhaps emphasizing one over the others depending on the context. These three approaches form the \textit{a priori} categories for our framework for thinking about and understanding the role of writing in physics labs.  

\subsection{\label{sec:labwriting}Writing in labs}
Most writing in the undergraduate physics curriculum takes place in lab classes. Some form of scientific writing is often included in learning goals of lab classes at all levels~\cite{KozminskiEtal2014, ZwicklFinkelsteinLewandowski2013, Dounas-FrazerEtal2018}, yet there is little research documenting why and how we incorporate writing specifically in lab classes. This paper represents an effort to begin to fill that gap, responding to calls from researchers for more attention to writing in labs~\cite{Lerner2007, StanleyLewandowski2016}.   

Typically, lab classes include both formative and summative types of writing in the form of lab notebooks and reports. Stanley and Lewandowski~\cite{StanleyLewandowski2016} conducted interviews with physics graduate students about their lab notebook practices and found that most graduate students did not receive formal instruction around keeping a lab notebook in their undergraduate lab courses or from advisors in graduate school. Most of the study participants reported eventually developing adequate lab notebook practices through informal experience in authentic research settings. The authors wonder if lab notebook documentation skills are explicit learning goals of lab courses, and if so, how these courses attempt to teach those skills. The framework we present in this paper broadly addresses possible goals for writing in lab classes, including documentation in lab notebooks. Some instructors have explored alternative structures or formats to the traditional lab notebook in order to teach students good record-keeping habits and facilitate the logistics of student-designed projects. For example, Eblen-Zayas~\cite{Eblen-Zayas2015} describes the transition to using Electronic Lab Notebooks (ELNs) and notes that they are particularly useful for instructors to track and guide student progress on multiweek projects. Students also reported that the ELNs were useful for organizing their ideas and data and facilitating collaboration. In Eblen-Zayas' example, the ELNs were graded and carried weight in the students' overall grade in order to emphasize the importance of keeping a lab notebook. Stanley and Lewandowski~\cite{StanleyLewandowski2018} provide additional recommendations for instruction around lab notebooks including having flexibility around the format and structure of notebook entries, clearly conveying to students that they should attend to \textit{context}, \textit{audience}, and \textit{purpose} in their notebook entries, and designing lab activities such that students have to rely on their own (or others') notebooks.    

In addition to lab notebooks, many physics lab classes employ lab reports, yet, in their traditional form, lab reports may have limited use. Traditional lab reports are known to be unsuccessful in promoting engagement in learning and quality writing, prescriptive of an artificial scientific process and information flow, time consuming to grade, and are often used by instructors without a clear purpose~\cite{Lane2014, AlaimoEtal2009, Haagen-Schuetzenhoefer2012}. Some alternative approaches to lab reports have been developed in order to address these shortcomings. The Science Writing Heuristic (SWH)~\cite{KeysEtal1999} is a tool that leverages the WTL approach, using the process of writing to help students make meaning of science content and practices. It consists of templates for instructors to design lab activities and templates to guide students in their writing and thinking about the activities. The SWH emphasizes the role of inquiry in science through questions that encourage ``deeper thinking and understanding about science concepts''~\cite{RuddGreenboweHand2001} and that guide students to make claims supported by evidence. It has been used in introductory chemistry labs in place of a traditional lab report, and improved students' motivation and attitudes toward science as well as their understanding of chemistry~\cite{RuddEtal2001}. 

Another alternative to a traditional lab report is the ``Letter Home'' assignment~\cite{Lane2014} where students write a letter (email) to a friend or family member in which they describe the lab activity and relay the results and interpretation of those results. The goal of this assignment, as it has been implemented in introductory~\cite{Lane2014} and upper-division~\cite{RameyThesis2018} physics lab classes, is to give students experience communicating physics to a non-physics audience, and has been documented to better promote student engagement and quality writing~\cite{Lane2014, RameyThesis2018}. Lastly, some science lab classes culminate with students writing a full scientific paper that mimics an authentic journal article, in an effort to help students join a professional discourse community by learning how to construct an argument supported by evidence, and communicate it through professional style and format~\cite{AlaimoEtal2009}. 

As part of enculturating students into a professional discourse community, some lab courses engage students in a peer review process around their proposals or reports, given that peer review is an essential element of the scientific process and that the act of revision is crucial for developing one's writing skills~\cite{Bean2011, AtkinsElliot2019}. Calibrated Peer Review (CPR) is an online system designed to help improve students' reading and writing skills in science, while simultaneously mitigating grading workload for instructors~\cite{Robinson2001}. Through the CPR system, students submit a writing assignment, evaluate three calibration examples (written by the instructor to be of low, medium, and high quality), evaluate three of their peers' assignments, then re-read and evaluate their own writing. Students receive a ``calibration score'' based on how well they evaluate the three calibration examples; the calibration scores of reviewers serve as a weighting factor in determining the final grade of each student's written assignment. CPR was created with the intention of helping students learn about a topic or content area through writing about it (WTL), improve their writing skills, and practice critiquing others' writing; it allows instructors to include writing assignments in large classes without having to spend large amounts of time grading them. 

While CPR has been used predominantly in large introductory lecture classes, there are some instances of implementation in lab classes. For example, Margerum \textit{et al}~\cite{MargerumEtal2007} report on an introductory chemistry lab in which they implemented three short WTL assignments through the CPR system---an essay on absorption and emission in the hydrogen atom as an introduction to the upcoming laboratory project, a pre-lab writing assignment on background information for the lab, and a post-lab formal lab report. The students in this study made progress toward meeting the learning objectives of the project, including improvement of their technical reading and writing skills. Wise and Kim~\cite{WiseKim2004} used CPR in a writing-intensive chemical engineering lab class, where students worked in groups on lab projects and then submitted individual executive summaries to accompany their lab reports. The students reported that the CPR process helped them improve their writing skills, as well as identify important aspects of their experiment. To justify to students the importance of writing an executive summary, the assignment began with the following statement: ``Presenting your work to managers and colleagues will be a part of your daily life when you go to industry, and how important it is cannot be overemphasized.'' Thus, the goals of the writing assignment were not only to improve students' writing skills and facilitate their learning of the content, but also to prepare them for a realistic professional writing practice. In this way, CPR can be used in lab classes to facilitate both WTL and WAP approaches to writing. 

Another example of peer review implementation in lab classes that foregrounds the WAP approach is the Journal of the Advanced Physics Laboratory Investigation (JAUPLI), an online student journal designed to help students experience an authentic double-confidential peer review process~\cite{JAUPLI}. Students submit articles about their advanced lab projects, which are then reviewed by anonymous students at other institutions and have a chance of being published in the online journal. Students who participated in JAUPLI reported that it helped them understand the scientific peer review process, improved their own understanding of their experiment, and improved their scientific writing skills~\cite{JAUPLI}. 

Regardless of the format or specific details of writing in lab classes, one common goal is that students engage in reflection, a process known to be important for learning the content and practices of science~\cite{MasonSingh2010b, MayEtkina2002, Dounas-FrazerReinholz2015}. The idea of reflection may be implied in the processes of experimentation we teach our students, or it may be explicitly included in instructions and framing to students (e.g., the SWH includes ``reflection'' as the last section of the report~\cite{RuddEtal2001}). In an advanced lab course that includes open-ended projects, Eblen-Zayas implemented separate metacognitive activities in the form of individual written reflections and a corresponding class discussion~\cite{Eblen-Zayas2016}. For the individual written reflections, students were given a series of prompts that guided them in thinking about what they did, how they dealt with problems they encountered, and what they plan to do moving forward. The activities were intended to normalize students' feelings of frustration throughout their open-ended project, and resulted in improved student enthusiasm for, and confidence in, doing experimental work. This literature on approaches to writing in science, and specifically lab classes, informs our framework of goals for writing in physics lab classes, with a special focus on upper-division labs that include a multiweek final project.

\section{\label{sec:methods}Methods}
\subsection{\label{sec:collection}Data collection}
The present study on the role of writing in physics labs takes place as part of a broader project on identifying effective practices for multiweek final projects in upper-division physics labs. In order to begin to define the breadth of goals for incorporating writing in physics labs, we conducted an interview study with four instructors of upper-division lab classes. That is, in order to answer our general research question about why physics lab instructors \textit{might} incorporate writing, we started with our four participants and asked why \textit{do} these instructors incorporate writing? The instructors come from a variety of institutional contexts: private and public, selective and inclusive, bachelors, masters, and doctoral degree granting institutions, including predominantly white, multi-racial, and Hispanic serving institutions. The four instructors were interviewed about their approach to, and goals for, writing in lab classes as part of their participation in the broader research project. Of the four, there are two white women and two white men. The results and analysis of this interview study are not meant to be generalizable or representative of all physics instructors; we report demographic information for both the institutions and the instructors to provide context for our work, and note that the framework that we have developed, based in part on the interview study, can be applied to a wide variety of physics lab classes in a variety of contexts. 

The four instructors in our interview study were all teaching advanced lab classes for physics majors that include a multiweek final project at the culmination of the term. The projects typically consisted of students working in groups to design and conduct their own experiment, with various amounts of scaffolding and guidance provided by the instructor up front and throughout the project. Each class incorporated multiple forms of writing, with all of them using lab notebooks and some of them including white papers, proposals, reflections, and reports. Additionally, three of the courses implemented some version of peer review where students reviewed each others' writing, responded to reviewer comments, and revised their papers accordingly. The writing in these classes was a mix of individual and group assignments.     

The interviews were semi-structured, lasted between 55 and 100 minutes, and were conducted by the first author via video conference. After having each instructor describe their course, specifically focusing on the details of the final project portion, the majority of the interview centered around the ways they incorporate writing in the final projects. For each type of writing assignment (e.g., lab notebooks, final reports), we asked them: \textit{Why do you incorporate this type of written communication during final projects? How do you frame it? What is the end goal or purpose? What role does this writing play in advancing the project? How do you grade it?} We also asked them specifics about implementation (e.g., \textit{Is it a collaborative or individual document? In what ways do students get feedback on their writing?}), and concluded the writing-related part of the interview by having them reflect on writing in the general physics curriculum and the role they think labs do (or should) play in teaching scientific writing.   

\subsection{\label{sec:analysis}Data analysis}
We recorded and transcribed the four interviews, and then coded each transcript for the instructors' goals or reasons for incorporating writing in the final projects in their classes. The coding analysis consisted of an iterative process of creating, refining, and applying codes that would answer our research question about the instructors' goals for writing. The codes were a combination of emergent and \textit{a priori}. We began with broad categories and ideas around writing present in the literature (Communication, WTL, and WAP, see Section \ref{sec:background-writing}) and then identified more specific goals cited by the instructors (emergent). We mapped these goals onto the broader categories, connecting the instructors' ideas to the \textit{a priori} categories from the literature. The interviews were open-ended and conducted before the \textit{a priori} categories were identified, such that the interview questions did not lead the instructors to talk about writing in a certain way. Through discussions among the research team, we iteratively refined the definitions and categorization of the codes. The final codebook is available in the online supplementary material~\footnote{See Supplemental Material at [URL will be inserted by publisher] for the codebook that resulted from the coding analysis of interviews and that was used as a foundation for creating the framework.}. Because we were looking for \textit{existence} of codes, and not prevalence or frequency, no inter-rater reliability was necessary for the coding analysis. Instead, in Section \ref{sec:framework} below where we define and discuss the resulting framework, we present an example of each goal from the data (in the form of interview quotes), providing our argument for why each particular goal exists and is distinct from other goals.      

\subsection{\label{sec:framework-methods}Development of the framework}
Upon developing the codebook and coding the four interviews, we expanded the codebook into a framework (see Fig. \ref{fig:framework} below) based on common ideas in the literature around writing in science, as well as our research and teaching experience in experimental physics. The resulting framework (addressing the question \textit{Why might instructors incorporate writing during multiweek final projects in advanced labs?}) is thus more generally applicable to physics labs, though its development was originally based on an interview study with four instructors (which addressed the question \textit{Why do these instructors incorporate writing during multiweek final projects in advanced labs?}). All of the goals for writing mentioned by the four instructors in our interview study are corroborated by the literature. In Section \ref{sec:framework}, we describe each element of the framework, providing an example from the data and a brief discussion of where and how that particular idea appears in the literature. There are also a few areas of overlap and additional goals not explicitly mentioned in the interviews that we included in the framework because they are commonly discussed in the writing in science and/or physics education literature. These additions are noted in Section \ref{sec:framework} below, and indicated in the figure by an asterisk. After we created and refined the content and structure of the framework, using the codebook as a foundation and corroborating and supplementing with literature, we conducted a general face validity check with physics education researchers and lab instructors external to the project. Their input led to further refinements of the labels and descriptions of the goals (e.g., using ``argumentation'' instead of ``persuasiveness'' and clarifying the distinction between ``cohesive narrative'' and ``synthesis'').   

\section{\label{sec:framework}Framework}
\begin{figure*}
\centering
\includegraphics[scale=0.55]{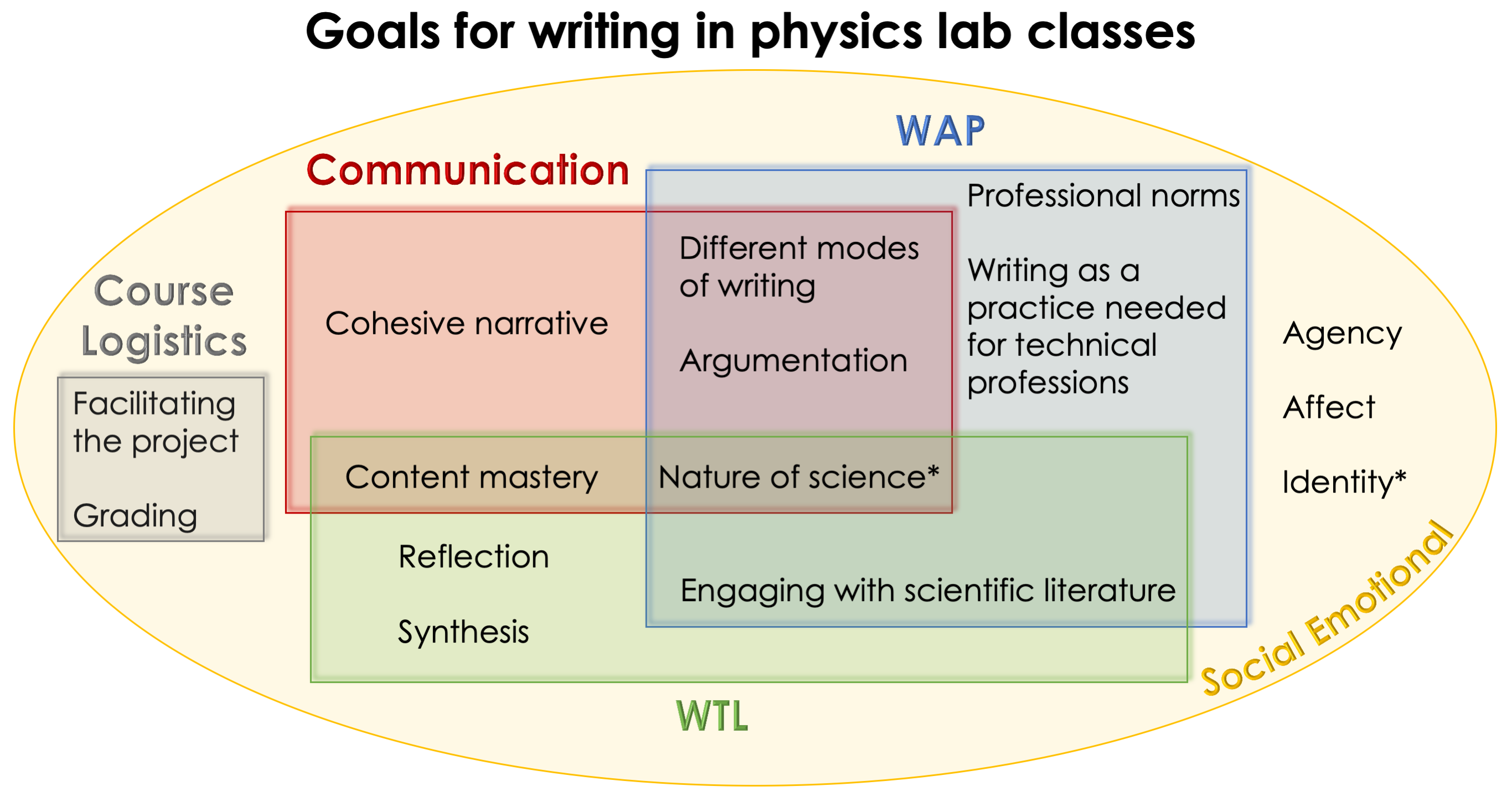}
\caption{Framework for thinking about and understanding the role of writing in physics lab classes. There are fifteen goals organized into five overlapping categories. WAP=Writing as professionalization. WTL=Writing to learn. *The Nature of Science and Identity goals did not appear in our data, but we identified them as important goals for the community of physics educators and thus added them to the framework.}\label{fig:framework}
\end{figure*}

The result of this research is the framework shown in Fig. \ref{fig:framework}. In this section, we define and describe each element of the framework. There are fifteen different possible goals for incorporating writing in physics labs, organized into five categories: \textit{Communication}, \textit{Writing as professionalization}, \textit{Writing to learn}, \textit{Course logistics}, and \textit{Social emotional}. The first three categories were \textit{a priori} (i.e., we identified these main ideas in the literature and used them to organize the list of goals that resulted from analyzing the data), while the latter two categories emerged from the data analysis process. As shown in Fig. \ref{fig:framework}, the categories are not mutually exclusive---they represent distinct, but interrelated, ideas and thus there are several instances of overlap between them. Further, the goals in the framework are of different grain sizes. They represent the breadth of possible goals for students that one might have when incorporating writing in a physics lab class, from a general understanding of writing as an important part of science, to specific skills like being able to write a cohesive narrative. 

\subsection{\label{sec:comm}Communication}
This category generally treats writing as communication, and focuses largely on the final product (rather than the process) of the writing. From the lens of communication, the primary purpose of writing is for the writer to demonstrate what they know or share what they did. The general idea of students being able to clearly communicate their work is a commonly cited objective for science classes generally~\cite{ReynoldsEtal2012} and physics lab classes more specifically~\cite{Eblen-Zayas2016, KozminskiEtal2014}. The goals in this category are about helping students develop general communication skills for the sake of communication, regardless of future career or profession. The \textit{Communication} category encompasses five different goals: \textit{Cohesive narrative}, \textit{Different modes of writing}, \textit{Argumentation}, \textit{Content mastery}, and \textit{Nature of science} (because this final goal lies at the intersection of many categories we include it in its own subsection, Section \ref{sec:NOS}). 

\subsubsection{\label{sec:cohesive}Cohesive narrative}
The underlying idea of this goal is that sections of a piece of writing are not separate and independent thoughts, but rather connect to one another to form a cohesive and consistent story. One goal of incorporating writing in physics labs is to help students be able to write a cohesive narrative, an important step in achieving clear and effective communication.

When talking about what they are looking for in students' lab reports, one instructor identifies a cohesive narrative as an important element: 
\begin{quote}
    \textit{``I think one of the biggest things for me is a coherent narrative...you would read it and feel like, okay, this could be written like a good scientific paper. It's coherent in the narrative. It ties the bigger picture in with what they're doing specifically, and then goes back to that bigger picture at the end. I think what a lot of students want to do is see each section as a separate thing. Like, here's my experiment. Here's the data I got. Let's talk analysis. You should be like, `Here's a graph. Let's talk about it and what it means. Or, here's how this might have tied in to the limitations of the equipment.' For me, it's really the fact that there's continuity and that the sections really work together to make the whole story. More than anything, that's what I think a good lab report should have.''---Instructor 1}
\end{quote}
Here, the instructor identifies coherence as an important element of a “good” lab report; their goal is for students to develop the skill of telling a cohesive story about their experiment. This instructor says elsewhere in the interview that they specifically coach students not to use the results section of their report as a ``data dump,” but rather to connect everything together to say what they did and what it means. Blakeslee~\cite{Blakeslee1997} documents a similar approach in an ethnographic case study of a graduate student learning to write a journal article with the guidance of a research advisor---in this case, the student had a tendency to focus first on technical descriptions and formatting such that early drafts of the paper looked like a journal article, but did not have a cohesive narrative. The role of the research advisor was thus to help the student learn how to communicate their results as a coherent story. As the instructor in our study indicates, this skill may be targeted and developed through students writing lab reports in advanced lab classes. 

\subsubsection{\label{sec:diffmodes-comm}Different modes of writing}
One goal for incorporating writing in final projects may be so that students can engage in different types of writing---writing with different purposes for different audiences. Many aspects of a given piece of writing, such as tone, organization, or level of detail shift depending on the context. Facility with navigating this context-dependence is one skill that we might want students to develop in a physics lab class. Indeed, practicing different types of writing that each require different skills~\cite{Blakeslee1997}, or being able to communicate to different audiences and write in different contexts are identified in science education literature as goals for students~\cite{Eblen-Zayas2016, ReynoldsMoskovitz2008}. The \textit{Different modes of writing} goal exists in both the \textit{Communication} and \textit{WAP} categories because you might imagine wanting students to practice the general communication skill of being able to write in different ways for different audiences, or you might want students to engage in different modes of writing \textit{because} that is what they will have to do if they go on to a career in science or engineering (writing proposals, reports, memos, etc.). In our interview study, instructors talked about the goal of having students engage in different modes of writing in ways that aligned with both the \textit{Communication} and \textit{WAP} categories. 

When talking about why they have students write proposals, one instructor said: 
\begin{quote}
   \textit{“Part of the goal is to...get them to write in a different context about experimental physics. Different contexts than just the lab reports.” ---Instructor 3} 
\end{quote}
This instructor has identified the general need for students to write about experimental physics in different contexts, where here we interpret ``contexts'' to mean different kinds of writing with different end goals (i.e., in a proposal, the writer is laying out a plan and convincing the reader of importance or feasibility, while in a report, the writer is describing what they have done and making conclusions based on evidence they have provided to readers). 

Instructors talked both about the general need for students to practice different types of writing (as illustrated in the above quote), but they also identified individual specific modes of writing that were important for students to engage in. Importantly, instructors who did this identified multiple individual modes of writing, from which we infer the goal of having students engage in different kinds of writing. We consider examples of these types of writing and separate them into \textit{formative} and \textit{summative} modes, based on the intended audience and purpose. Formative writing takes place as the project is happening, and is used in one way or another to help progress the project and/or to prepare for summative communication later on. The audience of formative writing could be the student themselves (now and in the future), their group members, or the instructor. Lab notebooks are the most common type of formative written communication in lab classes~\cite{StanleyLewandowski2016, Eblen-Zayas2015, KozminskiEtal2014}. When discussing the importance of lab notebooks, one instructor states:
\begin{quote}
    \textit{“It's like this is one of the really important modes of communication that you need to practice and you need to get good at. It's communication in this case, both with your...lab partners right now, and with yourself in the future.” ---Instructor 3}
\end{quote}
In describing this formative mode of communication as an important thing for students to practice, this instructor identifies the audiences to which this particular kind of writing is directed. 

In the context of final projects in labs, summative communication takes place at the culmination of the project (or one stage of the project). It may be written for peers, the instructor, or a general science audience with the primary purpose of having the student share what they did or what they learned. Lab reports are a common form of summative writing in lab classes; one instructor talks about the purpose and the audience of a final report:
\begin{quote}
    \textit{``I think that's the most important, to be able to really explain what you've done...I would say that...I'm the audience, but..maybe [also] anybody with an undergraduate physics degree.'' ---Instructor 4}
\end{quote}
Here, the instructor articulates the importance of a summative report by identifying the purpose (that students clearly explain what they did for their project) and audience (the instructor and general physics audience).

\subsubsection{\label{sec:argument-comm}Argumentation}
A goal for writing in final projects might be to help students learn how to develop a persuasive argument. Generally, the practice of argumentation involves convincing or persuading the reader, but it may also involve critique of your own or others' work by questioning models, assumptions, and claims. This goal falls under both the \textit{Communication} and \textit{WAP} categories because an instructor might find the general communication skill of being able to construct and deliver an argument to be important for their students to develop, but they may also tie the idea of argumentation directly to discipline- or profession-specific practices. 

One instructor, who has students write white papers and proposals as part of the process leading up to final projects, states:
\begin{quote}
    \textit{“We don't really focus on teaching writing skills as much as persuasive skills, making sure that the proposal has what it needs in there to convince people.'' ---Instructor 2}
\end{quote}
In the interview, this instructor made it clear that the goal of their class is \textit{not} to teach students the mechanics of writing (i.e., grammar and style), but rather to focus on the development of a persuasive argument. Argumentation, which involves making evidence-based conclusions and communicating them in a succinct and persuasive way, has been identified as a key learning outcome of physics labs~\cite{KozminskiEtal2014, ZwicklFinkelsteinLewandowski2013}. 

\subsubsection{\label{sec:content-comm}Content mastery}
A common goal of having students complete a writing assignment in any physics class is to have them learn physics content. The writing can facilitate the learning as well as be the medium through which students demonstrate their learning. In our data, the \textit{Content mastery} goal appeared only in the \textit{WTL} category, but we also include it in the \textit{Communication} category in the framework because part of content mastery requires that the learner demonstrates their mastery of the content, and this is often done through writing~\cite{Rivard1994}. This aspect of the \textit{Content mastery} goal is also connected to the \textit{Grading} goal---instructors might need students to demonstrate their understanding of the content through writing so that they can assign a grade. The instructors in our interview study did not talk about this communicative aspect of content mastery (aside from the specific connection to grading), and so we leave the example quote for the \textit{WTL} section below (Section \ref{sec:WTL}). 

\subsection{\label{sec:WAP}Writing as professionalization}

The \textit{WAP} category emphasizes the idea that writing is important because it is something you have to do as a scientist. Goals in this category focus on practices, norms, and skills that students will need to be proficient in if they become a professional physicist (or other related profession). While there is some overlap between the \textit{WAP} and \textit{Communication} categories, \textit{WAP} is more specific in that it focuses on the profession or discipline, whereas the \textit{Communication} goals are independent from a student’s major, career, or profession. Many recent reforms in physics lab classes have centered around the idea of WAP~\cite{MoskovitzKellogg2005}, using writing as a tool for developing students' sense of identity as a physicist and preparing them to participate in and contribute to the community of practicing physicists~\cite{IrvingSayre2014}. Part of this preparation involves communicating using ``forms authentic to the discipline''~\cite{KozminskiEtal2014, ZwicklFinkelsteinLewandowski2013}, or using authentic writing experiences to support authentic science experiences~\cite{Lerner2007}. There are six goals that fall under this umbrella: \textit{Different modes of writing}, \textit{Argumentation}, \textit{Professional norms}, \textit{Writing as a practice needed for technical professions}, \textit{Engaging with scientific literature}, and \textit{Nature of science}. We describe the first five here, and leave the discussion of the \textit{Nature of science} goal for Section \ref{sec:NOS} below).  

\subsubsection{\label{sec:diffmodes-wap}Different modes of writing}
As described above in Section \ref{sec:diffmodes-comm}, this goal is founded on the idea that it is important for students to be able to write about experimental physics in multiple contexts (i.e., writing with different purposes for different audiences). Whereas above we introduced the idea of \textit{Different modes of writing} under the general \textit{Communication} category, it may also be a specific goal of instructors to have students engage in different kinds of writing, both formal and informal, that mirror what they would have to do as professional physicists~\cite{AtkinsElliot2019}. That is, not only is it good to be able to communicate (in writing) different things to different audiences, but a practical skill for someone hoping to start a career as a physicist is to be able to write a report, a proposal, a memo, keep a lab notebook, etc. Indeed, JTUPP identifies being able to write for a variety of audiences as a learning goal that will promote career-readiness for physics undergraduate students~\cite{JTUPP2016}. In our interview study, when talking about how professionalism was an important part of writing in their class, one instructor said:
\begin{quote}
    \textit{``Professionalism for me is another one, and that can involve..the mechanics of writing and being cognizant of how you present a piece of communication. But I also want to get them understanding how you write for different types of audiences, what is an appropriate way to write for different audiences.'' ---Instructor 1}
\end{quote}
Here, the instructor situates \textit{Different modes of writing} within the professional practice of being a physicist. As a followup to this statement, the instructor then talks about the unique context of a final project where students are only focusing on one experiment (as opposed to a different lab each week) as a place where they want students to consider, \textit{``how would you present that single experiment in different ways?''} Being able to present information about an experiment in different ways to different people is a skill that students will need to practice as part of their training to become professional physicists. While this example illustrates generally the need to have students practice different kinds of writing as a part of the professionalization process, we also might imagine more specific examples of wanting students to practice particular modes of writing. Again, we identify formative and summative modes of writing. 

When talking about one common formative mode of written communication---lab notebooks---an instructor said:

\begin{quote}
    \textit{“I want them to get in the habit of taking notes about what they've done so that they have a record of what they've done, and so that they don't forget what they've done, and that they have a reference. And as a way to say, ``Okay, what did I accomplish today?" And have to look back and assess what it was that...they did. I think it's a good practice if you're going into research, to have a notebook of what you've done. I wanted to try to get them into that habit.'' ---Instructor 4}
\end{quote}
Here, the instructor identifies the audience (the student themselves) and the purpose (to have a record to look back on, and to facilitate self-reflection and assessment of what they have done so far) of the lab notebook. Likewise, Stanley and Lewandowski~\cite{StanleyLewandowski2018} identify the principles of \textit{context}, \textit{audience}, and \textit{timescale} as encapsulating the purpose of scientific documentation in lab notebooks. 

Another instructor identifies lab reports, a summative mode of written communication, as an example of a type of technical writing students should be familiar with. 
\begin{quote}
    \textit{“I feel like it is something that is not necessarily talked about and emphasized, but it's a skill that is realistic that they're going to have to do, is some type of technical writing. It's not always going to be a lab report format.'' ---Instructor 1}
\end{quote}
The instructor clarifies that lab reports are useful for having students practice summative written communication, but notes that in reality they may not actually be writing reports, but some other form of summative technical writing. This further explains the importance of helping students develop the ability to shift between different modes of writing, attending to the changing audience and purpose, a complex skill that is identified in science education as one that is important for students to practice~\cite{Gillen2006}. 

\subsubsection{\label{sec:argument-WAP}Argumentation}

As described above in Section \ref{sec:argument-comm}, one goal for including writing in final projects is to help students learn how to develop an argument and be persuasive. When this goal is specifically connected to preparing students for a profession or career---i.e., we want to help students learn to write persuasively \textit{because} being able to construct a persuasive argument is an important skill for a scientist---we include it in the \textit{WAP} category.

In talking about the role that persuasive writing plays in proposals, one instructor said:
\begin{quote}
    \textit{“I find as a scientist, it's a realistic thing you have to do. You very often have to justify why you want funding, why you want to be able to do this.'' ---Instructor 1}
\end{quote}

To an instructor assigning writing in a physics lab class, argumentation may be important for the sake of general communication skills (Section \ref{sec:argument-comm}), but it may also have discipline-specific meaning and importance. For example, in developing learning goals for an advanced laboratory course, physics education researchers and physics faculty at University of Colorado Boulder identified argumentation as a key aspect of the communication goals they had for students in their course~\cite{ZwicklFinkelsteinLewandowski2013}. In doing so, they narrow the general idea of argumentation to define specifically what the process of argumentation looks like in physics: not only convincing an audience of claims supported by evidence, but first justifying the appropriateness and accuracy of predictive models used to describe a reliable set of data. In this way, a goal of incorporating writing in a physics lab class might be to help students develop these argumentation skills specific to the discipline of (experimental) physics. Likewise, many researchers identify persuading skeptical audiences of the validity of your conclusions or interpretations as a common practice of scientists and thus something we need to help students develop~\cite{Gillen2006, MoskovitzKellogg2011, Blakeslee1997}. 

\subsubsection{\label{sec:professional-norms}Professional norms}

Another reason instructors might incorporate writing in lab classes is so that students can learn the discourse norms (i.e., rules or conventions) of the discipline, including style, format, tone, etc. If students continue on to a career in physics (or other related discipline), they will need to know how to write like a physicist---what goes in an abstract, how to write in a professional and scientific tone, and when to use or define jargon. We could imagine a goal of developing ``professionalism" to be extended more broadly to focus on how to speak and act in professional physics settings, but in this framework we focus only on the written communication element. We further narrow this goal to focus on specific norms or conventions that one must learn in order to contribute to a given professional community, as defined by community consensus, and do not include more general professional norms (e.g., physicists communicate their work through writing papers). These more general items are encompassed in other areas of the framework. 

One instructor articulates the professional norms goal when talking about how they want students to write final lab reports: 
\begin{quote}
    \textit{“Part of it is just generally...the mechanics and the formats of writing. If you're expected to write a technical report, you need headings...It's the language you choose to use. It's things like, okay, a graph needs to be readable and not be Excel defaults. It's stuff that, okay, if an employer hires you, they would say, `This is good work.''' ---Instructor 1}
\end{quote}
This professor articulates the goal of having their students learn and adopt the often unspoken rules regarding mechanics and format of scientific writing, a goal that is echoed by the science education community~\cite{Gillen2006, Blakeslee1997, KozminskiEtal2014, ZwicklFinkelsteinLewandowski2013}. The process of learning these norms could be supported by engaging with scientific literature (Section \ref{sec:lit}) or practicing different modes of writing (Sections \ref{sec:diffmodes-comm} and \ref{sec:diffmodes-wap}). In this way, the possible goals for writing outlined in this framework are distinct from, yet can interact with, one another. 

\subsubsection{\label{sec:practice}Writing as a practice needed for technical professions}
One reason instructors might include writing in the class is because written communication is important for a variety of technical professions. This goal is not as specific as the others in the \textit{WAP} category; rather than identifying a specific element of writing, this goal generally addresses the writing practices or experiences common in technical professions (e.g., experimental physics, engineering, science in general). Instructors of upper-division lab classes typically want to prepare students for skills they will need beyond their lab classes assuming they go on to a career in physics or other technical profession~\cite{JTUPP2016, ZwicklFinkelsteinLewandowski2013}. Instructors in our data who espouse this goal talk about writing as a practice, habit, or generally as something that scientists do (including something they personally do in their daily life), and they often speak about the importance that writing holds in the process of science or in the practice of technical professions. For example, one instructor talks about keeping a lab notebook as a scientific practice:
\begin{quote}
    \textit{“Partially it's going back to the realism of the process, it's a really important thing, because if you work in any technical environment, you need documentation of what you're doing.'' ---Instructor 1}
\end{quote}
Here, the instructor describes the written communication of a lab notebook as a realistic part of the process in a technical environment, and thus it is important that students have a chance to practice and experience keeping a notebook. This is aligned with work in PER that identifies the scientific practice of keeping a lab notebook as a possible learning goal of physics lab classes~\cite{Eblen-Zayas2015, StanleyLewandowski2016, StanleyLewandowski2018}. Lab notebooks are perhaps the most obvious form of writing recognized as a \textit{practice} or \textit{habit} that we might want students to develop, yet other forms of writing are also integral to the process of science.

Another instructor touches on the \textit{Writing as a practice needed for technical professions} goal when discussing why lab reports are an important aspect of the final projects in their class:
\begin{quote}
    \textit{“...it's to give them practice writing. Specifically scientific writing of a scientific report in...the unique context where it's something that they have proposed and pulled together and completed from beginning to end...I think...having a solid conclusion to the work... [is] a characteristic of scientific life. Right? I mean you've got these projects. You start them. Maybe they don't go exactly how you think they should go, but nonetheless, the progress, the scientific process, writing new grants, everything depends on you writing reports wherever you got to, and I think that skill is part of the point...Getting practice of doing, carrying out that skill is one of the main reasons why we do final reports.'' ---Instructor 3}
\end{quote}

This instructor talks about writing reports as a ``characteristic of scientific life," and that even when your experiment does not go as planned, you have to be able to summarize what happened and identify next steps in some form of summative writing. Given that many of the students in their class will propose and conduct an experiment for their final project and either not get results or not be able to complete the experiment as planned, it is important to this instructor that students are able to write a concluding report, an experience that mirrors the reality of scientific life. The goal of having students see and understand writing as a scientific practice inherent to the process of scientific knowledge generation is supported in the literature on scientific writing~\cite{MoskovitzKellogg2011}. This is also connected to the \textit{Nature of science} goal, which we discuss below in Section \ref{sec:NOS}. 

\subsubsection{\label{sec:lit}Engaging with scientific literature}
Part of writing as a scientist involves reading scientific literature. One goal for incorporating writing in labs might be to help students develop the skill of finding relevant papers, reading them, and situating their own work or project within a broader scientific community. 

One instructor connects this goal specifically to helping students develop skills they will need to be a physicist. In describing the process of writing a proposal for the final project, they say that it should be implemented: 
\begin{quote}
    \textit{“...in a way that forces them to engage with the literature and to actually read and do a little bit of lit review kinds of things, which is an important...skill for being a physicist.'' ---Instructor 3}
\end{quote}

In our data, this goal showed up only in the \textit{WAP} category, as instructors talked about how literature reviews are a realistic practice of scientists. However, not only is this practice an important part of the scientific writing process, but reading scientific literature may facilitate stronger understanding of the content or subject. Thus, in the framework this goal straddles the \textit{WAP} and \textit{WTL} categories (Section \ref{sec:lit-WTL}).  

The AAPT recommendations for laboratory learning goals suggest that as part of ``communicating physics'' students should be able to interpret and evaluate the work of others~\cite{KozminskiEtal2014}; in order to develop and practice this skill, students first need to learn how to engage with scientific literature~\cite{Gillen2006}. Moskovitz and Kellogg argue for the importance of students reading primary scientific communication (scientists presenting original research to other scientists) in order to learn how to communicate as scientists~\cite{MoskovitzKellogg2005}, since reading scientific articles is a specific practice of scientists important to the process of experimentation and generation of scientific knowledge~\cite{Lerner2007}.

\subsection{\label{sec:WTL}Writing to learn}
The \textit{WTL} category emphasizes writing as a tool for thinking and learning, focusing on the \textit{process} of writing more than the final product~\cite{Rivard1994, ReynoldsEtal2012}. Goals in this category focus on the idea that writing can be used to facilitate learning of the content or practices of experimental physics, and that writing requires ``frequent practice, effective feedback, and continual revision''~\cite{Lerner2007}.

\subsubsection{\label{sec:content}Content mastery}

 Some form of writing is often used in science classes to help students learn science content (e.g., short answer response questions on homeworks or exams, reflection questions, or tutorial worksheets)~\cite{HandWallaceYang2004, ReynoldsEtal2012}. Though conceptual understanding is not always the primary learning objective in physics lab classes, we might assign writing in order to, among other things, facilitate content mastery. For example, using the Science Writing Heuristic (SWH)~\cite{KeysEtal1999} as a replacement for a traditional lab report in introductory chemistry labs has been shown to improve both students' attitudes toward, and conceptual understanding of, chemistry~\cite{RuddGreenboweHand2001, RuddEtal2001}. In the scenario of students designing and conducting their own experiments in an advanced physics lab, a goal for a given writing assignment (e.g., proposal or lab notebook) might be to help students learn and become familiar with the particular topic area associated with their project.

We see an example of this goal in our data when one instructor details their goals for having students write final reports:
\begin{quote}
    \textit{“[The end goal of the report is] both documentation for what they’ve done and a physics background understanding of the purpose of the lab.'' ---Instructor 4}
\end{quote}

In the framework, this goal straddles the \textit{WTL} and \textit{Communication} categories because the process of writing may facilitate learning content, but the final product is how students communicate (to themselves, their peers, or the instructor) what they have learned. 

\subsubsection{\label{sec:reflection}Reflection}
Reflection is often identified in STEM education as a practice or skill that can be beneficial for supporting students' learning, problem-solving, enthusiasm, confidence, persistence, and epistemological views of science~\cite{Eblen-Zayas2016, ReynoldsEtal2012, MasonSingh2010b, MayEtkina2002, Dounas-FrazerReinholz2015, McInernyEtal2015}. A common way to have students practice and engage in reflection in any physics class is through writing. In the context of final projects in lab classes, this might mean using writing to help students reflect on their learning, what they do and do not understand, the progression of their project, or the process of science in general. This reflection could be realized through specific reflective writing prompts, or through the process of a more typical form of written communication like a lab notebook or report. However, Lippmann Kung and Linder caution that, at least in a laboratory setting, we must pay attention not to the \textit{amount} of metacognition students are engaging in, but to the instances of metacognition that allow students to transition into a sense-making mode~\cite{LippmannKungLinder2007}.   

In our data, one instructor identified reflection as an element of the practice of keeping a lab notebook:
\begin{quote}
    \textit{“The reflective bit as well, I think is something that we emphasize in the lab notes. It's not just data points written without comment. Right? It's I think maybe when I envision this reflective goal, often times it's a little bit more like reflecting on your own learning process and things like that. But I think actually again, writing down how you figure out what's going on in your experiment when you don't know what the next steps are yet or what the new versions of the experiment might look like. Your thought processes haven't converged yet. I think maintaining the lab notebook, being forced to put into writing what you think is going on at every step, I think that feeds into that goal of reflecting on the process of doing experimental physics.'' ---Instructor 3}
\end{quote}

Depending on the context and the type of writing, we can encourage students to engage in different kinds of reflection. Multiweek final projects may be particularly well-suited to encouraging students to reflect on the process of experimental physics, as this instructor has indicated. 

\subsubsection{\label{sec:synthesis}Synthesis}

The process of writing requires and facilitates synthesis of ideas. One goal of having students engage in writing could be to support them in being able to make sense of what their project means and connect ideas together. Within the \textit{WTL} category, this goal focuses specifically on the sense making aspect of synthesis; sitting down to write about an experiment you conducted requires you to synthesize the ideas and thus can facilitate your learning of content or process. This goal can be closely connected to being able to construct a \textit{Cohesive narrative} (see Section \ref{sec:cohesive}). They are distinct in that the \textit{Synthesis} goal is more about the process of synthesizing, while \textit{Cohesive narrative} is more about the presentation and communication of a narrative. The former is likely required in order to construct the latter. This is an example of how the elements of the framework may be concurrent with, or depend on, one another. 

One instructor touches on the idea of synthesis as they reflect generally about why they incorporate writing in their class:
\begin{quote}
    \textit{“For me, the big thing about writing is writing to make a point and to make a point correctly and coherently. Not everyone's going to write a lab report. Not everyone's going to write a paper. But you have to know how do you put all these pieces of evidence and all this background, how can that work together for you to draw conclusions? I find it as a tool hopefully to help them understand how to analyze the situation.'' ---Instructor 1}
\end{quote}

In line with the WTL approach, this instructor describes writing as a tool for analysis---the act of writing provides opportunity for students to practice drawing conclusions from evidence and synthesizing many disparate pieces. These sense making, analysis, or construction of meaning processes are often goals for students in physics laboratory classroom or professional environments~\cite{MoskovitzKellogg2011, ReynoldsEtal2012, Blakeslee1997}. 

\subsubsection{\label{sec:lit-WTL}Engaging with scientific literature}
Through reading scientific literature, students may learn physics content and/or ways of thinking. The \textit{Engaging with scientific literature} goal only appeared in our data in the \textit{WAP} category because instructors focused on needing to learn the professional practice of reading, and situating your work within, a body of scientific literature. In our framework, we include this goal in the overlap between the \textit{WAP} and \textit{WTL} categories because the process of engaging with scientific literature may also facilitate learning physics content. This goal is aligned with scholars who identify the pedagogical potential of scientific literature as well as the need to read good writing in order to improve one's writing skills~\cite{Gillen2006, Bean2011, AtkinsElliot2019}.

\subsection{\label{sec:log}Course logistics}
The \textit{Course logistics} category includes goals for incorporating writing that are related to how the class or the final project functions: \textit{Facilitating the project}, and \textit{Grading}.

\subsubsection{\label{sec:facilitating}Facilitating the project}
One common goal of incorporating various forms of writing is to help or encourage the students to plan and make progress on their project. A given type of writing may be necessary in order to move the project forward, i.e., without a thorough lab notebook, a project might not have much chance of succeeding. In a study of lab instructors' learning goals, Dounas-Frazer \textit{et al.}~\cite{Dounas-FrazerEtal2018} found that advanced optics and electronics lab classes often had learning goals related to written communication. One instructor in their study ``explicitly connected students' ability to keep good notes to their ability to iteratively improve their experiments''~\cite[p. 16]{Dounas-FrazerEtal2018}. 

One instructor in our interview study talked about the role of proposals for the final projects in their class:
\begin{quote}
    \textit{“...for the proposal, we want them to be prepared so that their term project can be successful. When they start the term project they have the materials, they know what they're doing, they have everything designed.'' ---Instructor 2}
\end{quote}

In addition to learning content and practices, a given writing assignment (like a proposal) may be a necessary milestone that helps students initiate and complete an experiment.  

\subsubsection{\label{sec:grading}Grading}
Writing is often included in lab classes because the instructor needs to know what the student did and/or learned so that they can assign a grade~\cite{Rivard1994}. In many classes (especially those with large enrollment), instructors do not have direct or frequent access to students' work throughout the course of a lab project, and must rely on various forms of written communication in order to evaluate the student's progress. One instructor articulates this goal when talking about why they assign final reports:
\begin{quote}
    \textit{“There's some other boring answer about me needing to have the information about how they're actually thinking about things in the end.'' ---Instructor 3}
\end{quote}

\subsection{\label{sec:socemo}Social emotional}
The \textit{Social emotional} category includes goals for writing related to students’ personal experiences in the social environment. The \textit{Social emotional} elements may be facilitated by, or help to facilitate, goals in other categories. For example, the process of reflection may help to facilitate a positive affective response toward experimental physics, or the development of a sense of identity as a physicist may help to facilitate content mastery. The \textit{Social emotional} category is ontologically different from all the rest in that it encompasses \textit{experiences} or \textit{feelings} that we might want students to have, while the other categories primarily include writing-related \textit{skills} we want them to develop. The writing-specific goals (in the \textit{Communication}, \textit{WAP}, and \textit{WTL} categories) all take place within the social environment of a classroom community. Thus, we consider the \textit{Social emotional} category to be connected to, or underlying all the rest. In Fig. \ref{fig:framework}, we can think about the yellow \textit{Social emotional} oval as existing on a different plane beneath the others. This category emerged from the data, but encompasses things that are broadly important to the physics education community and are cited as potential benefits or goals of writing in lab classes~\cite{RuddGreenboweHand2001, Eblen-Zayas2016, KozminskiEtal2014}. 

Our understanding and representation of this category are informed by a sociocultural perspective of learning wherein we consider context to be integral to learning~\cite{LaveWenger1991, Finkelstein2005}. Like any act of cognition, we thus view writing as situated in context~\cite{ReynoldsEtal2012, Blakeslee1997}, as a process and product that involves people working together to construct understanding and generate knowledge~\cite{RuddGreenboweHand2001} while learning and adopting community norms and practices~\cite{IrvingSayre2014}. There are three goals in the \textit{Social emotional} category---\textit{Affect}, \textit{Agency}, and \textit{Identity}---which overlap and interact with each of the other goals in the framework, while still standing alone as a distinct category of possible goals for students engaging in writing in a physics lab class. In this way, we consider the four categories (\textit{Communication}, \textit{WAP}, \textit{WTL}, \textit{Course Logistics}) to reside in a \textit{Social emotional} bath, while the \textit{Social emotional} goals inform, and are informed by, the other goals. We illustrate some of these connections below. 

\subsubsection{\label{sec:affect}Affect}
In general, a goal of many physics classes is to facilitate positive affective responses for students; we might use writing as a tool to guide this experience. In the context of students conducting final projects in a lab class, an instructor might structure and implement writing assignments such that they facilitate positive affective responses to the process of experimental physics. For example, one instructor comments on the importance of having students write final reports even when their experiment does not work out as planned:
\begin{quote}
    \textit{“...emotionally, I think it's probably important to feel like [the project] came to a conclusion and to actually have good feelings about what it is to be an experimentalist even if the experiment didn't necessarily go as they wanted it to.'' ---Instructor 3}
\end{quote}
Here, the instructor identifies the students' emotions (feelings about what it means to be an experimentalist) as important outcomes of the final project. They suggest that writing a final report can help students feel good about what they accomplished in the class. In this particular example, the affective goal is connected to students' identity---the instructor wants writing in the class to facilitate positive affective responses such that students feel good about being an experimentalist. In the framework, we include \textit{Identity} as a distinct goal (see Section \ref{sec:identity} below), noting that it may be coupled with other goals such as \textit{Affect}. While this example quote illustrates an affective goal connected to identity, they do not necessarily need to be coupled. We also might imagine an instructor assigning a given type of writing in order to help the students have fun doing their projects, regardless of how they may (or may not) identify as a physicist or experimentalist. Positive affect is often identified as a goal of lab classes~\cite{LewandowskiBoltonPollard2018}, and specific writing assignments have been shown to facilitate positive attitudinal shifts~\cite{RuddGreenboweHand2001, RuddEtal2001} and increased enthusiasm~\cite{Eblen-Zayas2016} among students. 

\subsubsection{\label{sec:agency}Agency}
A goal of many physics lab classes is for students to have ownership over their project or experiment. The experience of ownership can be empowering for students~\cite{Podolefsky2014} and benefit their motivation~\cite{Milner-Bolotin2001}, feelings of pride~\cite{Little2015}, and persistence in STEM~\cite{HanauerGrahamHatfull2016}. In a multiple case study, Dounas-Frazer, Stanley, and Lewandowski~\cite{Dounas-FrazerStanleyLewandowski2017} investigated how students came to feel ownership over their final projects in an advanced lab class. They identified five dimensions of ownership---student agency, instructor mentorship, peer collaboration, interest and value, and affective responses---and found that: ``(i) coupling division of labor with collective brainstorming can help balance student agency, instructor mentorship, and peer collaboration; (ii) initial student interest in the project topic is not always a necessary condition for student ownership of the project; and (iii) student ownership is characterized by a wide range of emotions that fluctuate in time as students alternate between extended periods of struggle and moments of success while working on their project''~\cite[p.18-19]{Dounas-FrazerStanleyLewandowski2017}. In our data, the idea of ownership came up generally but instructors more often talked specifically about the agency element. Given that agency is an important part of ownership, and that writing is one way students can exercise and demonstrate control over their own project, we include agency as a specific goal in our framework for understanding the role of writing in labs. During a multiweek final project, writing could facilitate the cycles of emotion that Dounas-Frazer, Stanley, and Lewandowski documented. 

While outlining the goals and benefits of having students write final lab reports, one instructor says that one thing they want students to get out of writing the reports is:
\begin{quote}
    \textit{“Specifically scientific writing of a scientific report in...the unique context where it's something that they have proposed and pulled together and completed from beginning to end.'' ---Instructor 3}
\end{quote}
This same instructor speaks about how agency is built into the structure of the final projects when they state:
\begin{quote}
    \textit{“Part of the structure of the final project is that they get to pick what they're doing. So [the proposals are] a framework for them to formally pick and formally communicate to me, to themselves, to their group mates, exactly what they want to do and how they want to do it.'' ---Instrucor 3}
\end{quote}

Together, these quotes illustrate how students can exercise and demonstrate agency through different forms of writing. Not only can students have agency over the design of the project and associated writing, but through writing students may have the opportunity to direct their own learning~\cite{Blakeslee1997}. 

\subsubsection{\label{sec:identity}Identity}
There is a rapidly growing body of research on the importance of supporting the development of students' science identity (e.g.,~\cite{AllieEtal2009, Hyater-AdamsEtal2018, LockHazari2016, LewisEtal2017}). A major goal of many advanced physics labs is to prepare students to do research~\cite{ZwicklFinkelsteinLewandowski2013}, which includes helping students develop a sense of identity as experimental physicists (or scientists more generally). The AAPT recommendations for undergraduate physics laboratory curricula suggest that ``Through laboratory work, students should gain the awareness that they are able to do science"~\cite{KozminskiEtal2014}. Writing is a specific element of laboratory work through which students may come to see that they can participate in, and contribute to, the physics community~\cite{QuanElby2016, IrvingSayre2014}. Though the \textit{Identity} goal only came up indirectly in our interviews with instructors (e.g., through talking about student affect, as shown in the example quote in Section \ref{sec:affect}), we include it in the framework as it plays an important role in physics lab classes, particularly in multiweek final projects in which students are designing and conducting their own experiments. Further, there is something very personal about writing (even the ``objective,'' technical writing of an experimental physicist), that makes it particularly well-suited as a site for identity development. 

In an account of the history of teaching writing in science, Lerner articulates the central role of identity in the writing of a scientific article:
\begin{quote}
    \textit{``The scientific article as a way of thinking about the process and communication of science is tightly wound to its authors' identities as scientists or would-be scientists. In other words, the key questions, methods for addressing those questions, and ways of situating those questions and answers within an ongoing body of research speak to the human act of science, not merely to a static document''}~\cite[p. 214]{Lerner2007}.
\end{quote}

This goal has strong connections to the overall \textit{WAP} category, and specific goals of \textit{Content mastery}, \textit{Reflection}, as well as \textit{Agency} and \textit{Affect} (as we have described above). The concept of writing as professionalization is very much connected to identity development---if students feel like they can be an experimentalist or they enjoy doing experimental physics, then they might be more likely to learn and take up the professional practices and skills identified in the \textit{WAP} category of the framework (Section \ref{sec:WAP}). On the other hand, experiencing and adopting professional norms and practices could help students feel like they belong as a member of the profession or can contribute to the community of practice~\cite{IrvingSayre2014, QuanElby2016}. \textit{Identity} may be especially related to \textit{Content mastery} because when a student feels like they understand the content of physics, they are more likely to see themselves as a physicist, and if a student has a sense of identity as a scientist, they may be more likely to engage with the content or feel confident that they can learn the content. The connections between \textit{Identity} and \textit{Reflection} may be similarly strong given that reflection can involve processing an experience and attending to your personal feelings about it~\cite{Dounas-FrazerReinholz2015}. Additionally, the process of reflection may facilitate content mastery, which in turn may inform a sense of disciplinary identity. In this way, identity development may exist as a specific goal of writing in labs, but also have strong interactions with other goals. Like the \textit{Social emotional} category as a whole, identity can inform, and be informed by, the other writing-skills oriented goals. 

\subsection{\label{sec:NOS}Nature of Science}
The \textit{Nature of science} goal exists in the overlap between the \textit{Communication}, \textit{WAP}, and \textit{WTL} categories, with strong connections to the \textit{Social emotional} category. Because the idea of the nature of science plays a unique role in this framework, we describe it here in its own section, and discuss how it connects several different goals, spanning multiple categories. 

Goals of physics lab classes often involve supporting the development of students' attitudes, expectations, or beliefs about the nature of experimental physics~\cite{ZwicklEtal2014, WilcoxLewandowski2016}. In our framework, we use the term Nature of Science (NOS) as shorthand for Nature of Experimental Physics to refer broadly to epistemological beliefs about the nature of knowledge (what does it mean to generate knowledge, know, or learn in the discipline of experimental physics) as well as expectations about the process and practice of experimental physics. NOS beliefs are foundational to everything that happens in a lab class and as such, the \textit{NOS} goal exists at the intersection of the three writing skills categories (\textit{Communication}, \textit{WAP}, \textit{WTL}) in our framework. One might implement writing in a lab class in order to help students see written communication as an important part of how scientific knowledge is generated (\textit{Communication}). Or, taking a \textit{WTL} approach, one might use writing as a tool to facilitate learning about content and practices of science, including ``disciplinary ways of knowing''~\cite{ReynoldsEtal2012}, the methods and process of science~\cite{MoskovitzKellogg2011}, or having students reflect on their own epistemic views about science~\cite{ReynoldsEtal2012}. Additionally, writing may be used to cultivate specific epistemological views that align with professional practice in the discipline (\textit{WAP}). 

There are a variety of ways writing may be used to support students' NOS beliefs in a lab class. One example is the SWH~\cite{KeysEtal1999}, which aims to help students see science as a process of constructing explanations by making connections and building on prior knowledge. The SWH template encourages these views about the nature of science through writing by emphasizing inquiry as fundamental to the process of scientific knowledge generation~\cite{RuddGreenboweHand2001}. Another approach is to assign written reflections (either as separate assignments or as part of a lab notebook entry) about the students' own attitudes toward, or beliefs around, the nature of experimentation~\cite{Eblen-Zayas2016}. Additionally, the way that we implement or frame writing can send consequential epistemological messages to students. Incorporating a WTL approach can help to shift from a ``knowledge telling'' to ``knowledge generating'' epistemology, thereby helping students see science \textit{not} as a collection of facts, but as a way of thinking or process of meaning making~\cite{Blakeslee1997, ReynoldsEtal2012}. 

There are strong connections between the \textit{NOS} goal and the \textit{Social emotional} category. In a case study analysis of undergraduate students' research experiences, Quan and Elby~\cite{QuanElby2016} documented shifts in students' NOS views (toward a more nuanced view of science in which novices are able to meaningfully participate) that were coupled to shifts in their self-efficacy. Likewise, there may be interplay between the \textit{NOS} goal and the \textit{Agency}, \textit{Affect}, or \textit{Identity} goals that may be realized through writing. For example, we might imagine that the experience of having agency over their own experiment may help students come to see experimental physics as a messy and iterative process of knowledge generation, characterized by cycles of frustration and excitement~\cite{Dounas-FrazerStanleyLewandowski2017} in which ``nothing works the first time''~\cite{Dounas-FrazerLewandowski2016}. 

Further, we can think of the \textit{NOS} goal, \textit{Social emotional} category, and writing skills categories (\textit{Communication, WAP, WTL}) as mutually informing or mediating one another, as discussed in the next section. 

\section{\label{sec:discussion}Discussion}
\subsection{\label{sec:connections}Connections between goals within and across categories}
We developed this framework primarily as a research tool to understand in depth the various goals we might have for students engaging in writing in a lab setting. As a scaffolding for organizing information, the framework provides a structure for the list of possible goals and organizes them into broader categories, allowing us to examine the interplay between goals within and across categories. For example, being able to engage with scientific literature and synthesize information, the processes of which can facilitate learning content and practices, may be in service to developing an argument and communicating it through a cohesive narrative. Further, attending to each of these specific aspects together may address a broader goal of supporting students' development of sophisticated views about the nature of science---the generation of scientific knowledge happens through a conversation among a community of scientists who synthesize information, make claims based on evidence, construct arguments and present them to one another, situating their argument among a body of scientific work. Thus, the distinct goals of \textit{Argumentation}, \textit{Engaging with scientific literature}, \textit{Synthesis}, \textit{Cohesive narrative}, and \textit{Nature of science} work together to paint a picture of the kind of writing we might want students to engage in in physics lab classes. 

We also see connections between the \textit{Course logistics} goals and other writing-skills focused goals, e.g., \textit{Grading} and \textit{Content mastery}. Writing is a primary way for instructors to find out what students are thinking or what they have learned; the process of writing can facilitate this learning (\textit{WTL}) and the final product can be used to communicate what they have learned (\textit{Communication}) for the purposes of practicing communication skills, sharing with peers, or receiving a grade in the course. 

The \textit{Nature of science} goal exists in the overlap between the \textit{Communication}, \textit{WTL}, and \textit{WAP} categories, and may play a unique role in connecting the writing skills oriented goals and the \textit{Social emotional} goals, which all mutually inform, or mediate, one another. For example, students views about the nature of science are informed by their experiences with, and attitudes towards, science (\textit{Affect}), which may also be informed by students' reflection on their own experiences of the process of experimental physics, or their own epistemic beliefs. If students have agency over an experiment, and exercise that agency, in part, through writing, they may reflect on that experience in a positive light and come to develop sophisticated views about the nature of science, which may help them to engage in the scientific process of constructing and communicating an argument. Further, teaching students about the nature and importance of argumentation in physics and giving them space to practice constructing and communicating arguments may help them come to view scientific knowledge as dialogic, tentative, and something that is constructed and advanced through written communication, which in turn may facilitate identity development. Developing a sense of identity as a physicist or having a positive affective experience may facilitate students' reflections about the content they are learning, the writing skills they are developing, or their own views about the nature of science. In this way, \textit{Nature of science} is the focal point that connects and mediates the other goals, which feels appropriate for lab classes where students learn what it means to \textit{do} physics, and thus supporting the development of students' views about the nature of experimental physics is of utmost importance. Considering the different approaches to writing and the interconnected nature of the goals helps us to make sense of the role that writing does, or should, play in physics lab classes. 

\subsection{Value of categories}
Ultimately, the boundaries we place between the goals and categories in the framework are artificial. As we have illustrated through a few examples, there may be overlap or simultaneity among goals, or the lines between them may be blurred in certain contexts. It is useful to delineate them and organize them into different categories so that we can have a common language with which to discuss what the goals mean, and to reflect on our own views of writing as instructors and researchers. For example, the way that we view writing impacts the way we tend to implement it, which in turn sends a message to students about the role of writing in science. If writing assignments are \textit{only} intended for the instructor (i.e., for evaluation purposes), this emphasizes the final product and a view of writing as demonstration or knowledge as telling~\cite{Bean2011}. The way that writing is employed and also evaluated in the classroom, will impact the way that students view and value it~\cite{Rivard1994, ReynoldsEtal2012}. If we are unaware of these views, or do not attend to them with our students through intentional choices in the classroom, students may develop views or habits counter to those we see as productive for their learning and participation in the discipline~\cite{BailyFinkelstein2015}.

\subsection{\label{sec:finalprojects}Benefits of final projects}
We developed this framework in the specific context of multiweek final projects in advanced lab classes. Though writing is typically incorporated in all types of lab classes (and these classes may address many of the goals described in the framework), project-based labs may be able to \textit{uniquely} facilitate the goals identified in the framework. We give a few examples here, noting that in a follow-up paper we will conduct case study analyses that will speak to the benefits (and limitations) of final projects, with respect to writing, in more depth. 

Stanley and Lewandowski~\cite{StanleyLewandowski2018} recommend that lab classes be structured in such a way so as to give real purpose to the lab notebooks, in order to facilitate authentic scientific documentation experiences for students. When students design and conduct their own experiments, often around a topic that the instructor is not familiar with, students have to rely on their own (and their group members') documentation practices in order to make progress on the project (\textit{Facilitating the project}). This also requires the students to engage in reflection throughout the project, evaluating at each step what the goal is, what they have done, the problems they have encountered, and how they plan to move forward (\textit{Reflection}). Engaging in this kind of reflection around a novel project with no clear answer or solution also encourages students to recognize the iterative nature of experimental physics (\textit{Nature of science}). An important element of final projects that can facilitate student engagement in reflection on their project as well as the nature of experimental physics is the long timescale---it can be difficult for students to do much reflection in one or two weeks. Working on the same project for several weeks provides time for this kind of deep thinking that can lead to refinement of the project. 

The longer timescale of final projects also provides ample time for students to revise their writing. Revision is a key element of helping students become better writers (and better scientists), and we need to teach them what it means to revise a piece of writing (i.e., not just editing for typos, but thinking deeply about organization of ideas and argument construction)~\cite{Bean2011}. The revision process can take a long time, and thus multiweek final projects are beneficial in this regard. Peer review can be a useful way to guide students through the revision process, and can be implemented in laboratory classroom settings in a variety of ways~\cite{MargerumEtal2007, WiseKim2004, JAUPLI}. Three of the four instructors we interviewed included a peer review process at some stage of the final projects. The process of peer review can help students to reflect on their experiment and their writing---reading and critiquing writing from peers can help students realize what they could have done differently on their projects or in their writing (\textit{Reflection}). The act of critiquing others' work can facilitate an understanding of what it means to construct and communicate an argument (\textit{Argumentation}) or to synthesize results (\textit{Synthesis}) and communicate them through a \textit{Cohesive narrative}. Additionally, depending on how the peer review process is structured in the context of the course, it can help students to understand what an authentic scientific writing process looks like and the role that revision plays~\cite{JAUPLI}. Thus, engaging in peer review can support students' views about the central role that writing plays in the generation of scientific knowledge (\textit{Writing as a practice needed for technical professions} and \textit{Nature of science}), facilitating identity development along the way (\textit{Identity}).  

An overall goal of advanced lab classes is often to prepare students to do research~\cite{ZwicklFinkelsteinLewandowski2013}. Compared to apprenticeship-style undergraduate research experiences, the environment of project-based labs can be beneficial because it allows students to have significant control over the experiment and the writing (\textit{Agency}), whereas in a real research project, the results of the experiment and the final written product matters for the professor, thus limiting the amount of control students can have~\cite{Blakeslee1997}. Additionally, in an ethnographic case study that explores the mentoring relationship between a graduate student and their advisor, Blakeslee documents the experience of a graduate student writing their first journal article~\cite{Blakeslee1997}. In this case, the student did not necessarily recognize the writing process as a learning experience itself, which impeded the ability of the advisor to guide the student through the process of synthesis, argument construction, and adoption of professional discourse norms. Blakeslee suggests that in making all kinds of learning goals explicit to students, ``students' learning can remain situated and embedded in activity while at the same time being more perceptible''~\cite[p.160]{Blakeslee1997}. Project-based labs have the benefit of making the learning explicit, and can convey to students that developing writing skills is a goal of the course along with learning laboratory skills like troubleshooting or experimental design.  

Lastly, many lab classes include a specific goal of having students engage in collaboration~\cite{JTUPP2016, KozminskiEtal2014}. We did not include collaboration in our framework because we do not see it as an end goal of engaging in writing, though they often exist in tandem. That is, engaging in writing may facilitate learning what it means to collaborate, but we do not assign writing in labs \textit{because} we want to teach collaboration. There are many other ways that we teach students about the importance of collaboration and how to do so effectively and equitably (e.g., structuring the course such that students must work in groups or rotate through assigned roles). Though learning about effective collaboration is not necessarily an end goal of having students engage in writing, we do see them as intimately connected. Future case study analyses will explore the role of collaboration in implementation of writing. 

\subsection{\label{sec:generalizability}Limitations and Generalizability}
The framework shown in Fig. \ref{fig:framework} is the result of a coding analysis of four interviews with instructors of advanced physics lab classes and synthesis of common ideas in literature on writing in science and/or lab classes. The goal of the coding analysis was to identify \textit{possible} goals for writing---because we were concerned with existence of codes and not prevalence, we did not count the frequencies of codes or look for patterns among the four interviews. Because the coding analysis of interview data was corroborated and supplemented by literature, the resulting framework is broadly applicable to physics lab classes. That is, the ideas represented by the framework are not unique to the course contexts of the four instructors we interviewed. Further, our overall research project focuses on the final project portion of advanced lab classes. Given the literature on writing in science and in lab classes in which our framework is grounded, we do not believe that the ideas represented by the framework are specific to project-based labs. Rather, student-designed multiweek projects may have unique affordances for addressing a variety of the goals for, and approaches to, writing presented in the framework.  

One limitation of this work is that we could have missed something that did not appear in the interview data or in the literature that we reviewed. In order to mitigate this issue, we presented the framework to physics education researchers and lab instructors external to this project in order to check for face validity and identify any obvious missing elements. After these discussions, the changes made to the framework were at the level of specific wording of goals and definitions. There were no additional goals or categories identified that were not already included in the framework.  

We feel confident that the resulting framework holds face validity with physicists who commonly teach lab classes. That said, we do not intend for this framework to be exhaustive, nor do we expect a single lab class to have each of these goals in mind when incorporating writing. Rather, it gives us a sense of possible goals for having students engage in writing in lab classes, what they might mean, and how they might interact with one another. 

\section{\label{sec:implications}Implications}
The framework is intended to be used by researchers or instructors as a tool to facilitate thinking about and understanding the role of writing in lab classes. Specifically, it can be used to: a) investigate or analyze the role that writing plays in different kinds of physics lab classes, b) inform future research questions, and c) inform pedagogical decisions or curricular design. 

\subsection{\label{sec:research}Implications for research}
There is a dearth of research on writing in physics lab classes. This paper begins to address that gap by providing a tool that researchers can use to investigate and understand the role of writing in lab classes. In a forthcoming paper, we will present case study analyses of project-based advanced lab classes, using our framework as a lens through which to view the role of writing in the lab classes. We anticipate that each class or instructor will target a different subset of the goals and that connections between goals may be more or less present depending on contexts and approaches. Using our framework as a research tool in this way will lead to deeper insights into the role that writing can play in project-based advanced labs. Other researchers could use this framework as a tool to study various lab classes. For example, it could be used to identify the existence and prevalence of different goals for writing across different kinds of lab classes (first year, beyond first year, project-based, verification labs, large enrollment, etc.) or different types of writing assignments (notebooks, proposals, reports, etc.).  

As a first step in beginning to address the dearth of research around writing in physics lab classes, this framework opens up several avenues for future research: How do physics instructors incorporate writing in lab classes in order to attend to some (or all) of the goals identified in the framework? How effective are the current practices around teaching scientific writing in lab classes? What are students' experiences with, and views around, writing in lab classes? Do students see the purpose of writing as being aligned with Communication, WTL, or WAP? How do different features or elements of lab classes impact students’ views around, and experiences of, scientific writing? We call on the physics education research community to begin to investigate these questions. 

\subsection{\label{sec:teaching}Implications for teaching}
The first step in designing, refining, or assessing a course is to define the learning goals. The framework we have presented here defines and describes \textit{possible} goals for implementing writing in lab classes. It may be a useful tool for instructors to help articulate or expand their thinking around the purpose of incorporating writing in lab classes. If an instructor wanted to focus on a particular goal, they may choose to structure writing assignments in a certain way. For example, if one wanted to emphasize WTL (or more specifically, reflection), they might emphasize the \textit{process} of writing in the timing, grading, and structure of a writing assignment by: including peer review, having students reflect on the revision process, having students turn in progressively more finalized drafts throughout the term, or grading students on the thoughtfulness of their revisions and not solely on the final written product.

Our future work will provide further resources for instructors by presenting case study analyses of what it may look like to implement writing in project-based advanced lab classes, in service of the goals defined here. 

\section{\label{sec:conclusions}Conclusions}
To create a framework for understanding the role of writing in physics lab classes, we conducted interviews with four advanced lab instructors, and supplemented the data with ideas from literature on writing in science. The resulting framework consists of fifteen possible goals that one might have for students when incorporating writing in a physics lab class, organized into five categories. The goals are distinct from, yet can interact with, one another.

Writing is an important part of science broadly, and experimental physics specifically. In the undergraduate physics curriculum, students encounter writing most frequently in lab classes, which often include ideas about communication or writing as explicit learning goals of the course~\cite{KozminskiEtal2014, ZwicklEtal2014, Eblen-Zayas2016}. We have begun to address the lack of research around writing in physics lab classes by investigating possible goals for writing. We see this as a first step toward understanding how to leverage writing to teach students physics content, engage students in practices and professional norms of experimental physics, help students develop clear communication skills, and support students' identity development in the domain of experimental physics. 

\begin{acknowledgments}
The authors would like to thank the instructors with whom we are partnering on this project, Dimitri Dounas-Frazer and Laura Ríos for prior work on this project and early drafts of the interview protocol, and members of the CU PER group for helpful feedback on the development of the framework. This work is supported by NSF grant DUE-1726045 and PHY-1734006. Viewpoints expressed here are those of the authors and do not reflect views of NSF. 
\end{acknowledgments}

\bibliography{Labs-Writing}

\end{document}